\documentclass{eptcs}
 \pdfoutput=1

\newcommand{\publicationstatus}{}

\usepackage{verbatim}

\usepackage{xspace}
\usepackage{latexsym}
\usepackage{amssymb}   %
\usepackage{ifthen}
\newboolean{commentsaon}         %
\setboolean{commentsaon}{true}
 \setboolean{commentsaon}{false}  %

\newboolean{commentson} %
\setboolean{commentson}{true}
 \setboolean{commentson}{false} %

\renewcommand{\comment}[1]
{\ifthenelse{\boolean{commentson}\AND\boolean{commentsaon}}
   {{\par\noindent\mbox{}{\footnotesize\blue[ *** #1 ]\par}\noindent\par}}{}}

\newcommand{\commenta}[1]
{\ifthenelse{\boolean{commentsaon}}
   {{\par\noindent\mbox{}{\small\color[rgb]{0, .5, 0}[ *** #1 ]\par}\noindent\par}}{}}

\usepackage[yyyymmdd]{datetime}

\renewcommand{\today}{2022-12-31}

\usepackage{color}
\newcommand\blue     {\color{blue}}

\newcommand{\dred}{\color[rgb]{0.7,0,0}}

\usepackage{hyperref}

\usepackage{bm}  %

\usepackage{underscore}  %
\usepackage{breakurl}

\newtheorem{theorem}{Theorem}
\newtheorem{corollary}[theorem]{Corollary}
\newtheorem{lemma}[theorem]{Lemma}
\newtheorem{proposition}[theorem]{Proposition}
\newtheorem{example}[theorem]{Example}
\newtheorem{definition}[theorem]{Definition}
\newtheorem{remark}[theorem]{Remark}

\newcommand*{\seq}[2][n]  {{#2_{1}, \allowbreak \ldots, \allowbreak #2_{#1}}}

\renewcommand*{\S}{{\ensuremath{S}}\xspace}

\newcommand*{\NN}{{\ensuremath{\mathbb{N}}}\xspace}

\newcommand*{\Var}{{\ensuremath{\it Var}}\xspace}
\newcommand*{\VarOut}{{\ensuremath{\it VarOut}}\xspace}
\newcommand*{\VarIn}{{\ensuremath{\it VarIn}}\xspace}
\newcommand*{\Ran}{{\ensuremath{\it Ran}}\xspace}
\newcommand*{\Dom}{{\ensuremath{\it Dom}}\xspace}

\newcommand*{\myunderscore}{\mbox{\tt\symbol{95}}}

\newcommand*{\mybackslash}{\ensuremath{\mbox{\tt \symbol{92}}}\xspace}
\newcommand*{\myeq}{\mathop{\doteq}}
\newcommand*{\MYEQ}{\mathop{\stackrel{..}=}}

\newcommand*{\nqueens}{{\sc nqueens}\xspace}

\title{A note on occur-check (extended report)}
\author{W{\l}odzimierz Drabent%
\vspace{1ex}
\\
\normalsize
\begin{tabular}{c}
    Institute of Computer Science,  Polish Academy of Sciences
\\
{\tt\normalsize\small
      drabent\,{\it at}\/\,ipipan\,{\it dot}\/\,waw\,{\it dot}\/\,pl}
\\  \today
\end{tabular}
\commenta{
}
}  %

\newcommand*\titlerunning{\today}
\newcommand*\authorrunning{\today}

\begin{document}
\maketitle

\begin{abstract}
  We weaken the notion of ``not subject to occur-check'' (NSTO), on which
  most known results on avoiding the occur-check in logic programming are
  based.  NSTO 
 means that unification is performed only on such pairs of atoms
  for which the occur-check never succeeds in {\em any} run of a
  nondeterministic unification algorithm. 
  Here we show that ``any run" can be weakened to ``some run''.  We present
  some related sufficient conditions 
  under which the occur-check may be safely omitted.
  We show examples for which the proposed approach provides more general
  results than the approaches based on well-moded and nicely moded programs
  (this includes cases to which the latter approaches are inapplicable).
  We additionally present a sufficient condition based on NSTO, working for
  arbitrary selection rules.
\\[.5ex]
\end{abstract}

\section{Introduction}

The programming language Prolog implements SLD-resolution employing an unsound
implementation of unification without the occur-check.  This usually creates
no problems in practice.  Programmers know that they do not need to care
about it, unless they deal with something unusual like checking a difference
list for emptiness.%
\footnote{%
  In the important Prolog textbook by Sterling and Shapiro
  \cite{Sterling-Shapiro-shorter}, the occur-check is mentioned
  (in the context of actual programs) only when discussing difference lists
  (p.\,298, p.\,300, on p.\,299 an error due to unsound Prolog unification
   is explained).  
  The textbook of Bratko \cite{Bratko.4ed} mentions the occur-check only
  once, informing that it is skipped in Prolog.
} %
Surprisingly, such attitude of programmers is often not justified by theory.
The known static criteria for occur-check freeness are applicable to restricted
classes of cases;
most of the approaches employing abstract interpretation are
restricted to the Prolog selection rule.  (See Section \ref{sec.final}
for an overview.)
Most of research on correctness of avoiding the occur-check is based
(explicitly or implicitly)
on the notion of NSTO
(not subject to occur-check) \cite{DBLP:conf/slp/DeransartFT91}.  It means
that unification is performed only on such pairs of atoms
for which the occur-check never succeeds in {\em any} run of a nondeterministic
unification algorithm.

In this paper we show that
 unification without the occur-check works
 correctly also for some cases which are not NSTO.
We  propose
a generalization of NSTO (Section \ref{sec.NSTO}).
We show that it is sufficient that the
occur-check does not succeed in
{\em some} run of the unification algorithm
for a given input, instead of {\em all} the runs (Section \ref{sec.WNSTO});
we call this
WNSTO (weak NSTO)
We discuss some related sufficient conditions %
for safely avoiding the occur-check.
They are applicable to some examples to which the former approaches are
inapplicable.  For other examples, a wider class of initial queries is dealt
with, or/and applying the proposed approach seems simpler than the former
ones. 

We additionally present a sufficient condition, based on NSTO 
(Section \ref{sec.tidy}), for safely
skipping the
occur-check under an arbitrary selection rule (and provide a detailed proof of
its correctness).
The condition generalizes that for nicely moded programs, by
Apt and Pellegrini \cite{AptP94-occur-check} and
Chadha and Plaisted \cite{ChadhaP94}.
In this paper we are mainly interested in syntactic sufficient conditions.
Approaches based on semantic analysis of programs, like abstract
interpretation, are left outside of the scope of this work.

The above mentioned sections are preceded by Section \ref{sec.preliminaries}
on preliminaries, and followed by Section \ref{sec.final}
containing an overview of related work and concluding remarks.
Two longer proofs and some additional results are placed in Appendices.
This paper is an extended version of \cite{drabent.iclp2021}.

\section{Preliminaries}
\label{sec.preliminaries}

We use the terminology, notation and many definitions  from \cite{apt-prolog}
(and reintroduce here only some of them).
The terminology is based on that of logic; in particular ``atom'' means an
atomic formula.

By an {\em expression} we mean a term, an atom, or a tuple of terms
or atoms.
An equation is a construct $s\myeq t$, where $s,t$ are expressions.
Given sequences of terms (or atoms) $s=\seq s$ and $t=\seq t$, 
the set $\{s_1\myeq t_1,\ldots,s_n\myeq t_n\}$ will be sometimes denoted by
$s\MYEQ t$.
A syntactic object (expression, equation, substitution, etc) is 
{\em linear} when no variable occurs in it more than once.
As in Prolog, each occurrence of \myunderscore\ in a syntactic object
will stand for a distinct variable.
Otherwise variable names begin with upper case letters.
$\Var(t)$ denotes the set of variables occurring in a syntactic object $t$.
We say that $s$ and $t$ are {\em variable disjoint} if 
$\Var(s)\cap\Var(t)=\emptyset$.

For a substitution $\theta = \{X_1/t_1,\ldots,X_n/t_n\}$, 
we define
$\Dom(\theta)=\{\seq X\}$,\, $\Ran(\theta)=\Var(\seq t)$, 
$\theta|S=\{X/t\in\theta \mid X\in S\,\}$
 (for a set $S$ of variables), and
$\theta|u= \theta| \Var(u)$
 (for an expression $u$).
   Note that $\Var(t\theta) \subseteq \Var(t)\cup\Ran(\theta)$
   (and thus $\Var(t\theta) \subseteq  \Var(t)\cup\Var(\theta)$),
 for any expression $t$ and substitution~$\theta$.

We employ
the Martelli-Montanari unification algorithm (MMA) 
(cf.\ \cite{apt-prolog}).
It unifies a (finite) set of equations, 
by iteratively applying one of the actions below
to an equation from the current set, until no action is applicable.  
The equation is chosen nondeterministically.
\[
\begin{array}{l@{\ }l@{\ \ } l l}
(1) & f(\seq s)\myeq f (\seq t)
    &  \longrightarrow &
    \parbox[t]{.37\textwidth}{replace by equations
      $s_1\myeq t_1,\ldots,s_n\myeq t_n$
    }
\\
(2) & f(\seq s)\myeq g(\seq[m] t)
      \makebox{ where $f\neq g$}
    &  \longrightarrow &  \mbox{halt with failure}
\\
(3) & X\mathop{\doteq} X  &  \longrightarrow
        & \mbox{delete the equation}
\\
(4) & t\mathop{\doteq}X 
  \mbox{ where $t$ is not a variable}
    &  \longrightarrow & \mbox{replace by }X\myeq t
\\
(5) &
    X\mathop{\doteq}t \mbox{ where }
    \begin{tabular}[t]{@{}l@{}}
      $X\not\in\Var(t)$ and  $X$ occurs elsewhere
    \end{tabular}
    &  \longrightarrow
    &
    \parbox[t]{.36\textwidth}{
      apply substitution $\{X/t\}$ to all other equations
    }
\\
(6) & X\mathop{\doteq}t \mbox{ where $X\in\Var(t)$ }
     \makebox[0pt][l]{and $X\neq t$}
      &  \longrightarrow & \mbox{halt with failure}
\end{array}
\]
By a {\em run} of MMA for an input $E$ we mean a maximal sequence
$E_0,\ldots,E_n$ of equation sets such that $E=E_0$ and
 $E_{i}$ is obtained from $E_{i-1}$ by 
one of the actions of the algorithm ($i=1,\ldots,n$).
Any run is finite, if a run does not end with failure then it results in an
equation set $E_n=\{X_1\mathop{\doteq}t_1,\ldots,X_k\mathop{\doteq}t_k\}$ in
a {\em solved form}  
($X_1,\ldots,X_k$ are distinct variables and none of them occurs in any term
$t_j$).
In such case, 
$\{X_1/t_1,\ldots,X_k/t_k\}$ is the obtained mgu (most general unifier).
See \cite{apt-prolog} for further details.

An mgu $\theta$ of $s,t$ is {\em relevant} if  $\Var(\theta)\subseteq\Var(s,t)$.
An equation set $E$ is said to be {\em NSTO}\/
if action (6) is not performed in any execution of MMA starting with $E$.
We often say ``unification of $s$ and $t$ is NSTO'' instead of 
``$\{s\myeq t\}$ is NSTO''.
In such case while unifying $s,t$
the occur check never succeeds, and thus can be skipped.

We need to generalize some definitions from \cite{apt-prolog}, 
in order not to be limited to LD-resolution.
We will say that {\em unification of $A$ and $H$ is available} in an
SLD-derivation (or SLD-tree)
for a program $P$,
if $A$ is the selected atom in a query of the derivation (tree), and $H$
is a standardized 
apart head of a clause from $P$,
such that $A$ and $H$ have the same predicate symbol.
(A more formal phrasing is {``equation set $\{A{\doteq}H\}$ is available''}.)
If all the unification{}s available in an SLD-derivation (SLD-tree) are NSTO
 then the derivation (tree) is  {\em occur-check free}.
We say that a program $P$ with a query $Q$ is {\em occur-check free}
if, under a given selection rule, the SLD-tree for $P$ with $Q$ 
is occur-check free.

  By an {\em answer} %
 of a program $P$ with a query $Q$
 we mean $Q\theta$ where $\theta$
 is an answer substitution for $P$ and $Q$.
(When it is necessary to distinguish between computed and correct answer
 substitutions, an answer related to a computed answer substitution will be
 called {\em computed answer}.)

Similarly to \cite{AptP94-occur-check}, we will employ modes.
This means dividing the argument positions of predicates into two groups,
by assigning a function $m_p\colon \{1,\ldots,n\}\to\{+,-\}$ (called a
{\em mode}) to each
predicate $p$ of the considered program (where $n$ is the arity of $p$).
A program with a mode for each predicate is called a {\em moded program}, 
and the collection of modes is called {\em moding}.
We follow the usual terminology, 
argument positions with $+$ assigned are called input, and those with $-$ are
called output.
We usually specify $m_p$ by writing $p(m_p(1),\ldots,m_p(n))$.
E.g.\ $p(+,-)$ states that the first argument of $p$ is input and the second
one is output.
Note that moding does not need to correspond to any intuitive notion of data
flow.

Given a moding,
we will write $p(s;t)$ to represent an atom $p(\seq t)$ and to state that $s$ is
the sequence of terms in its input positions, and $t$ the sequence of terms
in its output positions.
 An atom
 $p(s;t)$ is {\em input-output disjoint} if $\Var(s)\cap\Var(t)=\emptyset$.
Let us define $\VarIn(p(s;t))=\Var(s)$,  $\VarOut(p(s;t))=\Var(t)$.
The input (resp.\ output) positions of a query $Q$ are the input (output)
positions of the atoms of $Q$.
A query (or an atom) $Q$ is
  {\em input linear} (resp.\ {\em output linear})
if the sequence of the terms occurring in the
input (output) positions of $Q$ is linear.  
  We will refer to the following results. 
\begin{lemma}
\label{lemma.nsto}
  \begin{enumerate}
  \item
    \label{lemma.nsto.7.5part}
    Let $s$ and $t$ be sequences of terms, 
  such that the lengths of $s$ and $t$ are the same.
  If\, $\Var(s)\cap\Var(t)=\emptyset$ and $s$ 
    (or $t$) is linear
    then $s\MYEQ t$ is NSTO.
    (This is a special case of \cite[Lemma 7.5]{apt-prolog}.)

  \item 
    \label{lemma.nsto.7.14}

    Consider atoms $A$ and $H$.  If they are variable disjoint, one of them is
    input-output disjoint, one of them is input linear, and the other is output
    linear then $\{A\myeq H\}$ is NSTO  \cite[Lemma 7.14]{apt-prolog}.
  \end{enumerate}
\end{lemma}

  Obviously, $p(s)\myeq p(t)$ is NSTO iff $s\MYEQ t$ is NSTO.
Based on Lemma \ref{lemma.nsto}.\ref{lemma.nsto.7.14}, Apt and Pellegrini
\cite{AptP94-occur-check} introduced two sufficient conditions for
occur-check freeness. 
One ({\em well-moded} programs and queries
\DeclareRobustCommand{\myfragment}{Def.\linebreak[3]\,7.8, Corollary\,7.18}%
\cite[\myfragment]{apt-prolog})
implies that the input positions of the atoms selected in LD-trees are ground.
The other one, also proposed independently by 
Chadha and Plaisted \cite{ChadhaP94}, 
implies that the selected atoms are output-linear
({\em nicely-moded} programs and queries
\cite[Def.\,7.19, Corollary 7.25]{apt-prolog},
see also Section \ref{sec.tidy.condition}).

\section{NSTO and arbitrary selection rules}
\label{sec.tidy}
\subsection{Sufficient condition}
\label{sec.tidy.condition}
 
Here we propose a syntactic condition for occur-check freeness under
arbitrary selection rules. 
The next subsection contains examples.
We assume that the programs dealt with are moded.

\begin{definition}%
  Let  $Q=\seq A$  be a query.
We define a relation $\to_Q$ on $\{\seq A\}$.  Let
$A_i\to_Q A_j$ when a variable occurs in an output position of $A_i$ and in an
input position of~$A_j$.

  Query $Q$ is {\bf tidy} if it is output linear and $\to_Q$ is acyclic
  ($A_i\not\to_Q^+A_i$ for $i=1,\ldots,n$).

  Clause $H\gets Q$ is {\bf tidy} if $Q$ is tidy, and
  
\qquad  $H$ is input linear,  

\qquad
  no variable from an input position of $H$ occurs in an output position of $Q$.
\end{definition}

Note that each atom in a tidy query is input-output disjoint.
Also, if a query $Q$ is tidy then any permutation of $Q$ is tidy too.
A linear query is tidy under any moding.
This is a basic property of tidy clauses and queries:

\begin{lemma}
\label{lemma.tidy}
  Let $Q$ be a tidy query, and $C$ a tidy clause
  variable disjoint from $Q$.
  An SLD-resolvent $Q'$ of $Q$ and $C$ is tidy.
\end{lemma}
A proof is given in Appendix \ref{app.proof.tidy}.
Now our
 sufficient condition for occur-check freeness is:

\begin{corollary}
\label{cor.tidy}
\label{corollary.tidy}
  A tidy program with a tidy query is occur-check free, under any selection
  rule. 
\end{corollary}

\noindent
{\sc Proof }
By Lemma \ref{lemma.tidy}, 
in each SLD-derivation for a tidy program and query, 
each query is tidy.
Assume $A$ is an atom from a tidy query and $H$ is the head of a standardized
apart tidy clause.  As $A$ is input-output disjoint and output linear and $H$
is input linear, 
$\{A{\doteq}H\}$ is NSTO, by Lemma \ref{lemma.nsto}.
\hfill $\Box$.

\medskip

    Corollary \ref{cor.tidy} is a generalization of an earlier result.
    A {\bf nicely moded} query/program with input linear clause heads is a tidy
    query/program such that
     $A_i\to_Q A_j$ implies $i<j$ (for each involved relation $\to_Q$).
    Such programs/queries were shown to be occur-check free under LD-resolution
  \cite{ChadhaP94}, \cite{AptP94-occur-check}, \cite{apt-prolog}
  (the first reference uses different terminology).
  Apt and Luitjes \cite{AptL95.delays} generalized this
 for other selection rules (with the same restriction on $\to_Q$),
 however a proof seems unavailable.%
\footnote{
\cite{AptL95.delays} generalizes two results of
  \cite{AptP94-occur-check},  that for well moded and that for nicely moded
  programs.  The monograph
  \cite{apt-prolog} presents the first generalization,
  That related to Corollary  \ref{cor.tidy}
  is not even mentioned in ``Bibliographic Remarks''
  \cite[p.\,203]{apt-prolog}.  For its proof, the reader of \cite{AptL95.delays}
  is referred to an apparently unavailable master's thesis.
So Corollary \ref{cor.tidy} with its proof seems worth considering, despite it
may be viewed as
close to the result of \cite{AptL95.delays}.
} %
Finding modes under which a given program/query is tidy is outside of
the scope of this paper.

\subsection{Examples}
\label{sec.tidy.examples}

\noindent
 Apt and Pellegrini \cite[Section 7]{AptP94-occur-check}
state that
the approaches based on well-modedness or nice modedness are inapplicable
to some programs, and introduced a more sophisticated approach.
The programs are 
{\sc flatten} \cite[Program 15.2]{Sterling-Shapiro-shorter},
{\sc quicksort\myunderscore d{}l} \cite[Program 15.4]{Sterling-Shapiro-shorter},
and {\sc normalize} (\cite[Program 15.7]{Sterling-Shapiro-shorter});
they employ difference lists.
Here we focus on {\sc flatten}, which flattens a given list.
(We split arguments of the form $t\mybackslash u$ or $t\mbox{\small++}u$ 
into two argument positions; the argument positions of
 built-in predicates are moded as input.)
\pagebreak[3]
\[
\quad\ 
\begin{array}[t]{l}
\makebox[0pt]{\sc flatten:}
\\
    \%\,
    {\it flatten\_d l(X s,Y s,Zs) \mbox{ -- difference list }Y s\mybackslash Zs
    \mbox{ represents the flattened list }X s}
\\[.5ex]
{ \it
      flatten\_d l([X|X s],Y s, Zs) \gets
        flatten\_d l(X,Y s, Y s1),\
        flatten\_d l(X s,Y s1, Zs).
    }
    \\{ \it
      flatten\_d l(X,[X|X s], X s) \gets
         constant(X),\ X \mathrel{\mybackslash\mbox{\tt==}} [\,].
    }
    \\{ \it
      flatten\_d l([\,],X s, X s).
    }
\\[.5ex]
    {\it
      flatten(X s,Y s) \gets flatten\_d l(X s,Y s,[\,]).
    }
\end{array}
\]

\noindent
We see that, 
for the program to be tidy,
 the second and the third arguments of ${\it flatten\_d l}$ cannot be
both output (${\it Y s1}$ occurs in these positions in a clause body, which must
be output linear).  Also, the first and the second arguments of 
${\it flatten\_d l}$ cannot be both input, the same for the second and the
third one (as a clause head must be input linear).
The reader is encouraged to check that {\sc flatten} is tidy
under moding 
$M_1 = {\it flatten(+,-),\, flatten\_d l(+,-,+)}$,
and also under
$M_2 = {\it flatten(-,-),\, flatten\_d l(-,-,+)}$ and
$M_3 = {\it flatten(-,+),\, flatten\_d l(-,+,-)}$.
The relation $\to_Q$ 
(where $Q=A_1,A_2$ is the body of the first clause)
consists of one pair; for $M_1$ and $M_2$ it is $A_2\to_Q A_1$, 
for $M_3$ it is $A_1\to_Q A_2$.
The program remains tidy if the moding of {\it flatten} is
replaced by $(-,-)$ (in $M_1$ or $M_3$).

For any term $t$ and variable $R\not\in\Var(t)$, 
query $Q_0={\it flatten(t,R)}$ is tidy under $M_1$.
  (To be tidy under $M_2$ or  $M_3$, $Q_0$ has to be linear.)
By Corollary \ref{cor.tidy}, {\sc flatten} with $Q_0$ is occur-check free,
under any selection rule.

We only mention that  {\sc quicksort\myunderscore d{}l}
and {\sc normalize} are also tidy, and thus are occur-check free for a wide
class of queries.  
{\sc normalize} is similar to  {\sc flatten}, and is tidy for similar modings.
{\sc quicksort\myunderscore d{}l} is tidy 
for instance for modings\,
${\it quick sort(+,-)}$,\,\,${\it quick sort\_d l(+,-,+)}$, ${\it partition(+,+,-,-)}$,
\,and\,
${\it quick sort(-,+)}$, ${\it quick sort\_d l(-,+,-)}$, ${\it partition(-,+,+,+)}$.
Another example of a tidy program is {\sc derivative} (Example
\ref{ex.derivative} below).
 {\sloppy\par}

Surprisingly, it is not noticed in \cite{AptP94-occur-check} that
the approach based on nice modedness {\em is} applicable to
{\sc flatten} and {\sc normalize}.  
{\sc flatten} is nicely moded under $M_3$, so is the query $Q_0$, provided it is
  linear.  As the clause heads are input linear, it follows that {\sc flatten}
  with $Q_0$ is occur-check free for the Prolog selection rule.
Similarly, occur-check freeness can be shown for {\sc normalize}
  (as it is nicely moded under ${\it normalize\_d s(-,+,-)}$);
we skip further details.
In both cases the modes may be seen as not natural; what is understood as
input data appears in a position moded as output.
This may explain why the nice modedness of the programs was missed
in \cite{AptP94-occur-check}.

\section{Weakening NSTO}
\label{sec.NSTO}
The discussion on avoiding the occur-check in all the referred work
and in the previous section
is based on the notion of NSTO.
We show that NSTO is a too strong requirement.  
Unification without the occur-check produces correct results also for some
pairs of atoms which are not NSTO.
In such cases the algorithm may temporarily construct infinite terms, but 
eventually halt with failure.  An example of such pair is
$p(a, f(X), X )$, $p(b, Y, Y )$.  For this pair, some runs of MMA
 halt due to 
selecting $a\mathop{\doteq}b$, some other ones due to a successful occur-check.
Omitting the occur-check would result in failure on $a\mathop{\doteq}b$;  
this is a correct result.

NSTO requires that each run of MMA does not perform action (6).  In this
section we show that it is sufficient that there {\em exists} such run.
For this we need to introduce a precise description of the algorithm without
the occur-check, called MMA$^-$ (Section \ref{sec.MMA-}).
We define WNSTO, a weaker version of NSTO, and show that MMA$^-$ produces
correct results for expression pairs that are WNSTO (Section \ref{sec.WNSTO}).
Then (Section \ref{sec.nqueens})
we show an example of a program with a query which is not occur-check
free, but 
will be correctly executed without the occur-check, as all the atom pairs to
be unified are WNSTO.
In Section \ref{sec.sufficient.wnsto} (and Appendix \ref{app.weakly.tidy})
 we present
sufficient conditions for safely skipping the occur-check, based on WNSTO.

\subsection{An algorithm without the occur-check}
\label{sec.MMA-}
\noindent
By abuse of terminology, we will write
``unification algorithm without the occur-check'', despite such algorithm
does not correctly implement unification.
We would not consider any actual unification algorithm of Prolog,
this would require to deal with too many low level details.
See for instance the algorithm of
\cite[Section\,2]{DBLP:books/mit/AitKaci91}.
Instead, we use a more abstract algorithm, obtained from MMA.
We cannot simply drop the occur-check from MMA (by removing action (6) and 
the condition $X\not\in\Var(t)$ in action (5)).  The resulted algorithm 
may not terminate, as an equation $X\mathop{\doteq} t$ where $X\in\Var(t)$ can be selected
repetitively.

We obtain a reasonable algorithm in two steps. 
Before dropping the occur-check, MMA is made closer to
actual unification algorithms.  
The idea is to abandon action (5), except for $t$ being a variable.
(This action applies $\{X/t\}$ to all equations except one.)
Instead,when $t$ is not a variable,
$\{X/t\}$ is applied only when needed, and only to one occurrence of $X$.
This happens when the variable becomes the left hand side
of two equations $X\myeq t,\ X\myeq u$
(where $t,u$ are not variables, and $X\not\in\Var(t,u)$).
Then $X\myeq u$ is replaced by $t\myeq u$.
For technical reasons (termination), we require that $t$ is not larger than $u$.

Now, dropping the occur-check results in an algorithm presented by Colmerauer
\cite{Colmerauer1982}.
Without loss of generality we assume that we deal with unification of terms.
Let $|t|$ be the number of occurrences in $t$ of variables and
function symbols, including constants.

\pagebreak[3]
\begin{definition}[{\rm\cite{Colmerauer1982}}]
{\bf MMA$\!^{\boldsymbol-}$} (MMA without the occur-check) is obtained from 
MMA by
\[
\begin{tabular}{l@{ }l}
(a) & removing action (6), and 
\\
(b) & replacing action (5) by
\\[1ex]
\multicolumn{2}{l}
{%
  \begin{tabular}{l@{ }l}
    (5a) &
    \parbox[t]{.75\textwidth}{
    $X\mathop{\doteq}Y$, where $X,Y$ are distinct variables and $X$ occurs elsewhere
\\
      \mbox{}\qquad$\longrightarrow$ \qquad 
      \begin{tabular}[t]{l}
      apply substitution $\{X/Y\}$ to all other equations, \\
      \end{tabular}
    }
  \end{tabular}%
}
\\[2.5ex]
\multicolumn{2}{l}
{%
  \begin{tabular}{l@{ }l}
    (5b) &
    \parbox[t]{.85\textwidth}
{
      $X\mathop{\doteq}t$, $X\mathop{\doteq}u$ 
where $t,u$ are distinct non-variable terms;
    let $\{s_1,s_2\}=\{t,u\}$ and $|s_1|\leq|s_2|$
       \\
       \mbox{}\qquad$\longrightarrow$ \qquad
      \begin{tabular}[t]{l}
      replace  $X\mathop{\doteq}s_2$  by $s_1\mathop{\doteq}s_2$
      \end{tabular}
    }
  \end{tabular}%
}
\end{tabular}
\]
A set $E$ of equations is in a {\bf semi-solved form} if $E$ is
$\{X_1\mathop{\doteq}t_1,\ldots,X_n\mathop{\doteq}t_n\}$, where $X_1,\ldots,X_n$ are distinct,
each $X_i$ is distinct from $t_i$,
and if a $t_i$ is a variable then $X_i$ occurs only once in $E$
(for $i=1,...,n$).
\end{definition}

\noindent
Note that
 $E$ is in a solved form
 if, additionally, $X_i\not\in\Var(t_j)$ for $1\leq i,j\leq n$.
Note also that inability of performing any action of MMA$^-$ means that the
equation set is in a semi-solved form.

Let us first discuss termination of MMA$^-$.
Let $k>1$ be an integer greater than the arity
of each function symbol appearing in the equation set $E$
which is the input of the algorithm. 
We define a function $||\ ||$ assigning natural numbers to
equation sets and equations:
\[
\begin{array}{c}
  || \{\, t_1{\doteq}u_1,\ldots, t_n{\doteq}u_n \,\} || = 
  \displaystyle\sum_{i=1}^n ||t_i{\doteq}u_i||,  \qquad \mbox{where}
\qquad
  ||t{\doteq}u|| = k^{\mbox{\,\small$\max(|t|,|u|)$}}.
\end{array}
\]
Assume that an action of MMA$^-$ is applied to a set $E$ of equations,
resulting in $E'$.
Then $||E||=||E'||$ if the action is (4), (5a), or (5b), and 
$||E||>||E'||$ if it is (3).
By the lemma below, $||E||>||E'||$ for action (1).

\begin{lemma}
\label{lemma.norm}
For any terms (or atoms) $s=f(\seq s),\ t=f(\seq t)$,
\ \ 
$
 ||  s{\doteq} t||>\displaystyle \sum_{i=1}^n ||s_i{\doteq}t_i||.  
$

\end{lemma}
{\sc Proof%
\footnote{%
  Function $||\, ||$ is proposed and
  the lemma is stated without proof in \cite{Colmerauer1982}.
  There, however, $k$ is
  the maximal arity of symbols from $E$, which does not
    make sense when it is 0 or 1. 
    The lemma also holds (with a slightly longer proof) for $k$ being the
    maximal number out of 2 and the arities of the symbols.
} %
 } %
The inequality obviously holds for $n=0$.  Let $n>0$, and
$l,r$ be, respectively, the left and the right hand side of the inequality.
Without loss of generality, assume that $|s|\geq |t|$.
Now
$l = k\cdot k^{|s_1|}\cdots k^{|s_n|} \geq k\cdot k^{|t_1|}\cdots k^{|t_n|}$.
Hence
$l/k\geq k^{|u|}$ for any $u\in\{\seq s, \seq t \}$.
Thus
$l/k\geq ||s_i{\doteq}t_i ||$ for $i=1,\ldots,n$, and then
$n\cdot l/k\geq r$.  As $n/k<1$, we obtain $l>r$.\,%
\hfill
$\Box$

 Let $f_{45a}(E)$ be  the number of those equations from $E$
to which action (4) or (5a) applies.
Let $f_{5b}(E)$ be  the number of equations of the form $X{\doteq}u$ in $E$,
where $u$ is not a variable.
Note that applying (4) or (5a) decreases $f_{45a}(E)$, 
and applying (5b) decreases $f_{5b}(E)$ without changing 
$f_{45a}(E)$, 

Now consider the lexicographic ordering $\prec_3$ on $\NN^3$ (cf. for
instance \cite[p.\,33]{apt-prolog}).
If $E'$ is obtained from $E$ by applying one action of the algorithm,  
it holds that
\[
 (||E'||, f_{45a}(E'), f_{5b}(E')) \prec_3  (||E||, f_{45a}(E), f_{5b}(E)).
\]
Thus MMA$^-$ terminates, as $\prec_3$ is well-founded.  Summarizing:

\begin{lemma}
MMA$^-$ terminates for any input set $E$ of equations.
It either halts with failure (action (2))
or produces an equation set in a semi-solved form.  
\end{lemma}

In discussing further properties of the algorithm,
we will consider possibly infinite terms ({\bf i-terms}) over the given
alphabet.  (See \cite{DBLP:journals/tcs/Courcelle83} for a formal definition.)
We require that 
the set of variables occurring in an i-term is finite. 
The corresponding generalization of the
notion of substitution is called {\bf i-substitution}.

\begin{definition}
\label{def.i-}
A substitution (respectively i-substitution) $\theta$ is a {\bf solution}
({\bf i-solution}) of an equation $t\mathop{\doteq}u$ if
$t\theta=u\theta$;
\,\,$\theta$ is a {\em solution} ({\em i-solution}) of a set $E$ of equations, 
if $\theta$ is a {solution} ({i-solution}) of each equation from $E$.
Two sets of equations are {\bf equivalent} (respectively {\bf i-equivalent})
  if they have the same set of solutions (i-solutions).
\end{definition}

\begin{lemma}
\label{lemma.equivalent}  
If an action of MMA or of MMA$^-$ replaces an equation set $E$ by $E'$ 
then $E,E'$ are i-equivalent (and hence equivalent).

\end{lemma}
{\sc Proof }
For any i-substitution $\theta$,  $f(\seq s)\theta= f (\seq t)\theta$ \,iff\,
$s_i\theta=t_i\theta$ for all $i\in\{1,\ldots,n\}$.  Thus the claim holds for action (1).
For actions (3), (4) the claim is obvious; for (2), (6) it is void.
Actions (5) and (5a) replace an equation set
$E=E_X\cup E_1$ by $E'=E_X\cup E_1\{X/t\}$, where $E_X=\{X\mathop{\doteq}t\}$.
Consider an i-solution $\theta$ of $E_X$.
So $X\theta = t\theta$.
Hence $(V\{X/t\})\theta = V\theta$ for any variable $V$,
and thus $u\{X/t\}\theta=u\theta$ for any expression $u$.
So $\theta$ is a solution of $E_1$ iff  $\theta$ is a solution of 
$E_1\{X/t\}$.
For (5b), 
equivalence of $\{X{\doteq}t,X{\doteq}u\}\cup E_1$ and $\{X{\doteq}t,t{\doteq}u\}\cup E_1$ 
follows immediately from Def.\ \ref{def.i-}
\hfill$\Box$

\begin{lemma}
\label{lemma.semi.solved}  
Any set of equations $E$ in a semi-solved form has an i-solution.
\end{lemma}
{\sc Proof }
If an equation of the form $X{\doteq}Y$ occurs in $E$ then
$E$ has an i-solution iff $E\setminus\{X{\doteq}Y\}$ has an i-solution
(as such $X$ occurs in $E$ only once), 
Hence we can assume that $E$ does not contain any equation of this form.
Now the result follows from Th.\,4.3.1 of 
\cite{DBLP:journals/tcs/Courcelle83}.

For a direct proof, consider  $E=\{X_1\doteq t_1,\ldots,X_n\doteq t_n \}$,
where no $t_i$ is a variable.
Let $\theta=\{X_1/t_1,\ldots,\linebreak[3]X_n/ t_n \}$.   
Let $i\in\{1,\ldots,n\}$.  Then $X_i\theta^{j+1}=t_i\theta^{j}$ for any
$j\in\NN$.  Note that no variable from $\Dom(\theta)$ occurs in  $t_i\theta^{j}$
at depths%
\footnote{
   We say that a symbol $f$ occurs in a term $f(\seq[k]u)$ ($k\geq0$)
   at depth $1$, and that $f$ occurs in $g(\seq[k]u)$ at depth $j>1$ if $f$
   occurs in
   some $u_i$ ($1\leq i\leq k$) at depth $j-1$.  E.g. $X$ occurs in $f(X,a)$ at
   depth $2$.

   We say that $f(\seq[k]u)$ and $f(\seq[k]{u'})$ ($k\geq0$) are identical at
   depths $\leq1$ and that $f(\seq[k]u)$ and $f(\seq[k]{u'})$ are identical 
   at depths $\leq j$ (for a $j>1$) provided that each $u_i$ and $u'_i$ are
   identical at depths $\leq j-1$.
}
$\leq j+1$ (by induction on $j$, as $\theta$ replaces each $X\in\Dom(\theta)$
by a non-variable term).
Hence $t_i\theta^{j}$ and $t_i\theta^{j+1}$ are identical at depths $\leq j+1$.
Let $u_i$ be the i-term which, for each $j\in\NN$, is identical with
$t_i\theta^{j}$ at depths $\leq j+1$. 
We show that the i-substitution $\varphi=\{X_1/u_1,\ldots,X_n/u_n\}$
is an i-solution of $E$
{\sloppy\par}

Obviously, $X_i\varphi$ is identical with $t_i\theta^{j}$ at depths $\leq j+1$.
Note that $\theta^j = \{X_1/t_1\theta^{j-1},\ldots,X_n/t_n\theta^{j-1}\}$.
So for each $i$, terms $X_i\theta^j=t_1\theta^{j-1}$ and $X_i\varphi$ 
are identical at depths $\leq j$.
Hence %
$u\theta^j$ and $u\varphi$ are identical at depths $\leq j+1$, for any
non-variable term $u$. 
By transitivity (taking $u=t_i$),  we obtain that
$X_i\varphi$ and  $t_i\varphi$ are identical at depths $\leq j+1$
(for any $j$).
Thus $X_i\varphi = t_i\varphi$.
\hfill $\Box$

\comment{Explain \NN}
\medskip

In remains to discuss the results of MMA$^-$.  
Note that if $E$ and $E'$ are i-equivalent then
$E$, $E'$ are equivalent$\!$. %
Consider a run $R$ of  MMA$^-$ starting from an equation set $E$.
If $R$ halts with failure (due to action (2)) then, by
Lemma \ref{lemma.equivalent}, $E$ has no solutions (is not unifiable).
If it halts with equation set $E'$ in semi-solved form, then
by Lemma \ref{lemma.equivalent}, $E$ is unifiable iff $E'$ is.
So applying MMA to $E'$, which boils down to applying actions (5) and (6),
either halts with failure, or produces a solved form $E''$, representing
an mgu of $E$.  

Prolog does not perform the occur-check, and treats the semi-solved form
as the result of unification.  
Prolog implementations present the result to the user in various ways.
For instance the answer to query
$g(X,X)=g(Y,f(Y))$ is displayed as $X {=} Y,\, Y {=} f(Y)$ by SWI, and as
$X{=} f(f(f(\ldots))),\, Y{=} f(f(f(\ldots)))$ by SICStus
(predicate =/2 is defined by a unit clause ${=}(Z,Z)$).

\subsection{WNSTO}
\label{sec.WNSTO}

\noindent
Let us say that a run of MMA is {\em occur-check free} if the run does not
perform action (6). (In other words, 
no equation $X=t$ is selected where $X\in\Var(t)$ and $X\neq t$;
simply -- the occur-check does not succeed in the run).
An equation set $E$ is {\bf WNSTO} (weakly NSTO) when there exists an
occur-check free run of MMA for $E$.
When $E$ is $s\myeq t$ we also say that the unification of $s$ and $t$ is WNSTO.
A program $P$ with a query $Q$ is {\bf weakly occur-check free} if, under a
given selection rule, 
all the unification{}s available in the SLD-tree for $P$ with $Q$ are WNSTO.
A run of MMA$^-$ on an equation set
 $E$ is {\bf correct} if it produces correct results
i.e.\ the run halts with failure if $E$ is not unifiable, and produces a
unifiable equation set $E'$ in a semi-solved form otherwise.
The latter means that
iteratively applying action (5) to $E'$ produces an mgu
of $E$, in a form of an equation set in a solved form.
We say that  MMA$^-$ is {\bf sound} for $E$
if all the runs of  MMA$^-$ on $E$ are correct.
\medskip
WNSTO is sufficient for the unification without the occur-check to work
correctly:
\begin{theorem}
\label{th.MMAminus}
  Consider an equation set $E$.  Assume that there exists an occur-check free
run of MMA on $E$.
Then  MMA$^-$ is sound for $E$. 
\end{theorem}
{\sc Proof }
Let $R_1$ be an occur-check free run of MMA on $E$, and $R_2$ be a run
of MMA$^-$ on $E$.  
We show that $R_2$ is correct.
Let $S$ be the set of the i-solutions of $E$, and thus of every equation set
$E'$ appearing in $R_1$ or $R_2$ (by Lemma \ref{lemma.equivalent}).

If $R_1$ succeeds then $S$ contains unifiers of $E$, and of every $E'$
appearing in $R_2$.  Hence action (2) is not performed in $R_2$,
and $R_2$ halts with success producing a unifiable equation set $E_2$ in a
semi-solved form.

If $R_1$ halts with failure then the last performed action is (2), 
thus $S=\emptyset$.
This implies that $R_2$ does not produce a semi-solved form
(by Lemma \ref{lemma.semi.solved}).  Hence $R_2$ terminates with failure,
due to action (2).
\hfill$\Box$

\medskip
It immediately follows that a weakly occur-check free program $P$ with a
query $Q$ can be correctly
executed without the occur check:

\begin{corollary}
\label{corollary.main}
Assume a selection rule.
If a program $P$ with a query $Q$ is weakly occur-check free
then algorithm MMA$^-$ is sound
for each unification available in the SLD-tree for $P$ with $Q$.
\end{corollary}

We conclude this section with a technical lemma.

\begin{lemma}
  \label{lemma.iteration}
  Let $E_1\cup E_2$ be an equation set and $E_1$ be WNSTO.

  If $E_1$ is not unifiable then  $E_1\cup E_2$ is WNSTO.

  If  $\theta_1$ is an mgu of $E_1$, and $E_2\theta_1$ is WNSTO
  then  $E_1\cup E_2$ is WNSTO.

\end{lemma}
For a proof see Appendix \ref{app.proof.lemma.iteration}.

\begin{corollary}
\label{corollary.iteration}
    Consider a moding and atoms $p(s;t)$ and $p(s';t')$, where 
    $s\MYEQ s'$ is WNSTO.

    If  $s\MYEQ s'$ is not unifiable then 
    $p(s;t)\myeq p(s';t')$ is WNSTO.

    If $\theta$ is an mgu of $s\MYEQ s'$, and
    $(t\MYEQ t')\theta$ is WNSTO
    then  $p(s;t)\myeq p(s';t')$ is WNSTO.

\end{corollary}
{\sc Proof }
Equation $p(s;t)\myeq p(s';t')$ is WNSTO iff equation set 
$s\MYEQ s'\cup t\MYEQ t'$ is WNSTO.
Now Lemma \ref{lemma.iteration} applies.
\hfill$\Box$

\subsection{Example -- a weakly occur-check free program}
\label{sec.nqueens}
The core fragment of the $n$ queens program \cite{Fruehwirth91} will 
be now used as an example. We call it \nqueens, see 
\cite{drabent.nqueens.tplp}
 for explanations.
 \[
\renewcommand*{_}{\myunderscore}
\begin{array}{l@{\hspace{5em}}l}
    p q s(0,_,_,_).     &         
        (\refstepcounter{equation}\theequation\label{clause1})        \\
\it
    p q s(s(I),Cs,Us,[_|D s]) \ \gets\
            p q s(I,Cs,[_|Us],D s),\
            p q(s(I),Cs,Us,D s).   &
               \refstepcounter{equation} (\theequation\label{clause2})%
    \\[1ex]
    p q(I,[I|_],[I|_],[I|_]).     &
               \refstepcounter{equation} (\theequation\label{clause3})%
          \\
\it
    p q(I,[_|Cs],[_|Us],[_|D s]) \ \gets\
            p q(I,Cs,Us,D s).  &
               \refstepcounter{equation} (\theequation\label{clause4})%
\end{array}
\hspace*{-6em}
 \]

\noindent
The program works on non-ground data.
A typical initial query is 
$Q_{\rm in}={\it p q s}(n,q_0,\myunderscore,\myunderscore)$, where
$q_0$ is a list of distinct variables, and $n$ a natural number represented
as $s^i(0)$.

We now show that the standard syntactic approaches to deal with avoiding the
occur-check are inapplicable to \nqueens.
Under no moding the program is
well-moded with $Q_{\rm in}$ because 
there exist non-ground answers for $Q_{\rm in}$ 
(cf.\ \cite[Corollary 7.11]{apt-prolog}).
To be tidy,
at most one position of ${\it p q}$ is input (as 
(\ref{clause3}) must be input linear). 
Thus at least three positions of ${\it p q s}$ have to be output
(as a variable from an input position of the head
of (\ref{clause2}) cannot appear in an output position of body atom
   ${\it p q( s(I), Cs, Us, D s)}$).  This makes the body not output linear,
contradiction. 
Due to the same reasons, the sufficient condition for nicely moded programs
\cite{AptP94-occur-check}
is inapplicable.

It can be shown that \nqueens with $Q_{\rm in}$ is occur-check free under any
selection rule, by showing that in all SLD-derivations each atom in each query
is linear \cite{drabent.occur-check.report}. 
This is however rather tedious.
The program is not occur-check free for some non linear queries, for instance
for $A_{\rm S TO}={\it p q}(m, L, [L|\myunderscore],\myunderscore)$ (where $m$ is ground).
This is because unifying $A_{\rm S TO}$ with the unit clause  (\ref{clause3})
is not NSTO. 

We now show that \nqueens can be correctly executed without the occur-check,
for a wider class of initial queries.
Let us say that a query $Q$ is {\em 1-ground}
\phantomsection
\label{place.1-ground}%
if the first argument of the predicate symbol in each atom of $Q$ is ground.
We show that:

\begin{proposition}
 \nqueens is weakly occur-check free, under any selection
rule, for any 1-ground query.  
\end{proposition}

\noindent
 {\sc Proof }
Note first that, in each SLD-derivation for \nqueens and a 1-ground query,
each query is 1-ground.
Let $A=p(\seq[4]s)$ be a 1-ground atom, and
 $H={\it p q}(I, [I|\myunderscore], [I|\myunderscore], [I|\myunderscore])$
be the head of (\ref{clause3}), standardized apart.
Equation $s_1\myeq I$
  is NSTO and $\theta=\{I/s_1\}$ is its mgu.
Let $s=s_2,s_3,s_4$ and   %
$t=[I|\myunderscore], [I|\myunderscore], [I|\myunderscore]$.
As $s_1$ is ground, $t\theta$ is linear.
Thus $s\theta\MYEQ t\theta$ is NSTO by Lemma \ref{lemma.nsto}.
Hence by Lemma  \ref{lemma.iteration}, $s_1,s\mathop{\MYEQ} I,t$ is WNSTO.
So $A\myeq H$ is WNSTO.
The cases of the remaining clause heads of \nqueens are obvious, as the
heads are linear.
For another proofs, see Examples \ref{ex.nqueens.free},
\ref{ex.nqueens.free.syntactic}.
\hfill $\Box$
{\sloppy\par}

\medskip

By Corollary \ref{corollary.main},
\nqueens with with any 1-ground query is correctly executed
without the occur-check, under any selection rule.

\nqueens may be considered a somehow unusual program.  However
similar issues appear with rather typical programs dealing with ground data.
Assume, for instance, that data items from a ground data structure are to be
copied into two data structures.  Program
\[
\phantomsection
\label{place.use2}
\mbox{\sc use2:}
\quad
{\it
p( [X|X s], \, f(X,X s1),\,  [g(X,\myunderscore)|X s2]
 ) \gets p(X s,X s1,X s2).
\qquad
p([\,],a,[\,]).
}
\]
provides a concise example.
The elements of a list (the first argument of $p$) are copied into a term
$f(\ldots)$ and a list.
The program is not occur-check free for some 1-ground
queries (e.g.\ for $p([1], f(Y,Z),[Y|T])$).%
\footnote{{\sc use2}
    is well-moded under $p(+,-,-)$, but the approach for well-moded programs
    does not apply, as the clause head is not output linear.
 In contrast to \nqueens, {\sc use2} can be treated as tidy, or nicely moded.
 The program is tidy under any moding with at most one position $+$.
 Hence it is occur-check free for tidy queries
 (this includes all linear queries).

} %
Similarly as for \nqueens, it can be shown that
it is weakly occur-check free for all such queries.
For a quick proof see Ex.\,\ref{ex.nqueens.free} or 
\ref{ex.nqueens.free.syntactic}.

\subsection{Sufficient conditions for WNSTO}
\label{sec.sufficient.wnsto}
Now we discuss sufficient conditions for safely avoiding the occur-check
due to WNSTO.
We assume that the programs dealt with are moded.

We say that a selection rule is {\bf compatible with moding}
(for a program $P$ with a query $Q$)
if (i) the input positions are ground in each selected atom
in the SLD-tree for $P$ with $Q$
(\cite{AptL95.delays} calls this ``delay declarations imply the moding''), and
(ii) some atom is selected in every nonempty query.
Note that (i) may imply that the selection rule is partial,
in the sense that there exist nonempty queries in which no atom is selected.
This is called {floundering} (or {deadlock}).

An atom $A$ is {\bf weakly linear} if any variable $X$ which occurs more than
once in $A$ occurs in an input position of $A$.
(In other words, grounding the variables in the input positions of $A$
results in a linear atom.)
\begin{lemma}
\label{lemma.wnsto}
Consider variable disjoint atoms $A$ and $H$, such that the input positions
of $A$ are ground, and $H$ is weakly linear.  The unification of $A$ and $H$
is WNSTO.
\end{lemma}
{\sc Proof }
Let $A=p(s;t)$, where $s$ is ground, and $H=p(s';t')$.  Equation set
$s\MYEQ s'$ is NSTO (by Lemma \ref{lemma.nsto}).
Assume that $s\MYEQ s'$ is unifiable and that $\theta$ is an mgu of $s\MYEQ s'$.
Thus $X\theta$ is ground for each variable $X\in\Var(s')$.
Hence $t'\theta$ is linear, and $(t\MYEQ t')\theta$ is NSTO 
(by Lemma \ref{lemma.nsto}). 
Now by Corollary \ref{corollary.iteration},
$A\myeq H$ is WNSTO.
   \hfill$\Box$

\medskip
It immediately follows:
\begin{corollary}
\label{corollary.weakly..free}
  Let $P$ be a program in which each clause head is weakly linear.
  If the selection rule is compatible with moding then $P$ (with any query)
  is weakly occur-check free.  
\end{corollary}

\vspace{-1\medskipamount}
\begin{example}\rm
\label{ex.nqueens.free}

The heads of the clauses of 
\nqueens are weakly linear 
under moding ${\it p q s}(-,-,-,-)$, ${\it p q}(+,-,-,-)$.
By Corollary \ref{corollary.weakly..free}, the program
(with any query) is weakly occur-check free under
any selection rule compatible with moding.
Consider a query $Q$ which is 1-ground
(cf.\ Section \ref{sec.nqueens}, p.\,\pageref{place.1-ground}).
A simple check shows that in any SLD-derivation for \nqueens and $Q$ all
queries are 1-ground.  So 
each selection rule is compatible with moding (for \nqueens with $Q$).
Thus \nqueens with any 1-ground query is weakly occur-check free
for any selection rule.
 \ %
The same reasoning applies to {\sc use2}, with $p(+,-,-)$.
\end{example}

Now we provide a syntactic sufficient condition for a program to
be weakly occur-check free.
It employs a generalized notion of moding, in which some argument positions 
may be neither $+$ (input) nor $-$ (output),
to such positions we assign $\bot$ (neutral).
We will call it {\em\bf 3-moding} when it is necessary to distinguish it from
a standard moding.
We write $p(s;t;u)$ to represent an atom $p(\seq t)$ and to state that
$s$ (respectively $t$, $u$) is the sequence of terms in its $+$ ($-$,
$\bot$) positions.
The idea is to distinguish (as $+$ or $-$) some argument positions which,
roughly speaking, deal with ground data.
A syntactic sufficient condition will
imply for LD-derivations that in each selected atom the input positions are
ground.

By a {\bf well-3-moded} program (or query) we mean one which 
becomes well-moded \cite[Def.\,7.8]{apt-prolog}
after removing the $\bot$ argument positions.
For a direct definition, let a {\em defining occurrence} of a variable $V$ in
a clause $C = H{\gets} Q$ be an occurrence of $V$ in an input position of
H, or in an output position in $Q$.
Now $C$ is {\em well-3-moded} 
when each variable $V$ in an output position of $H$ has its
defining occurrence in $C$, and each occurrence of a $V$ in an
input position in $Q$ is preceded by a defining occurrence of $V$
in another atom of $C$
\cite{Dra87}.
A query $Q$ is well-3-moded when clause $p\gets Q$ is.
An equivalent definition can be obtained by an obvious adaptation of 
\cite[Def.\,7.8]{apt-prolog}.
Note that 
in a well-3-moded query the input positions of the first atom are ground.
Also, any query with its input positions ground is well-3-moded.

We will use the fact that
well-3-moded programs/queries inherit the main properties of well-moded ones.

\begin{lemma}
\label{lemma.well-3-moded}
  Let a program $P$ and a query $Q$ be well-3-moded. %
  \begin{enumerate}
  \item 
    \label{lemma.well-3-moded.queries}
    All queries in SLD-derivations of $P$ with $Q$ are well-3-moded. %

\item 
\label{lemma.well-3-moded.ground}
    If $Q\theta$ is an answer for $P$ with $Q$ then the input and output
    positions of $Q\theta$ are ground.

     \item
       \label{lemma.well-3-moded.Prolog}
       For $P$ with $Q$
       the Prolog selection rule is compatible with moding.

     \item
       \label{lemma.well-3-moded.nooutput}
       If no argument position is moded as output then 
       each selection rule is compatible with moding for $P$ with $Q$.

  \end{enumerate}
\end{lemma}

\noindent
{\sc Proof} \ 
For \ref{lemma.well-3-moded.queries}
it is 
  sufficient to show that any SLD-resolvent $R$ of a well-3-moded query
  $Q$ and a well-3-moded clause $C$ is well-3-moded.
Let $Q'$ (respectively $C'$ or $R'$) be $Q$ ($C$, $R$) with the $\bot$
arguments dropped.  So $Q',C'$ are well-moded, and $R'$ is their resolvent.
By Lemma 7.9 of \cite{apt-prolog}, $R'$ is well-moded.  Hence $R$ is 
well-3-moded.

For case \ref{lemma.well-3-moded.ground}, consider a successful SLD-derivation
$D$ for $P$ and $Q$ with a computed answer $Q\sigma$, of which $Q\theta$ is an
instance.  By dropping all the $\bot$ arguments from the queries and 
clauses of $D$, we obtain a successful derivation $D'$ for a
well-moded program and the well-moded query $Q'$.  Its answer  $Q'\theta$
is ground by \cite[Corollary 7.11]{apt-prolog}.
Thus the non $\bot$ positions of $Q\sigma$ and hence of $Q\theta$ are ground.

Claim \ref{lemma.well-3-moded.Prolog} follows from
 \ref{lemma.well-3-moded.queries},
as the input positions of the first atom of a well-3-moded query are ground.  
Under the 3-moding from claim \ref{lemma.well-3-moded.nooutput},
all the input positions in a well-3-moded query are ground.
So claim \ref{lemma.well-3-moded.nooutput} follows from
\ref{lemma.well-3-moded.queries}.
    \hfill $\Box$

\medskip
From the Lemma and Corollary \ref{corollary.weakly..free}
we immediately obtain:

\begin{corollary}
\label{cor.well-3-moded}
  Let $P$ be a well-3-moded program in which 
  each clause head in $P$ is weakly linear.
  Let  $Q$ be a well-3-moded query.  Then

  $P$ with $Q$  is weakly occur-check free under the Prolog selection rule;
  
  if    no argument position is moded as output then 
    $P$ with $Q$ is weakly occur-check free under any selection rule.

\end{corollary}

\begin{example}\rm
\label{ex.nqueens.free.syntactic}
Programs
\nqueens and {\sc use2} are well-3-moded under 
${\it p q s}(+,\bot,\bot,\bot)$, ${\it p q}(+,\bot,\bot,\bot)$, and
$p(+,\bot,\bot)$; so is any 1-ground query.  Their clause heads are weakly
linear. 
Thus by Corollary \ref{cor.well-3-moded},
 the programs are weakly occur-check free for 1-ground
queries, under any selection rule.
So we obtained by syntactic means
the results of Ex.\,\ref{ex.nqueens.free}.

\end{example}

\begin{example}\rm
\label{ex.derivative}

\noindent
Apt and Pellegrini \cite{AptP94-occur-check} use program {\sc derivative} 
\cite[Program 3.30]{Sterling-Shapiro-shorter}
as an example
for combining the approaches for well-moded and for nicely moded programs.  
Here we choose three representative clauses of the program
(infix operators ${\uparrow},{*}$ are used).
\[
\mbox{\sc derivative}\colon \qquad
\begin{array}[t]{l}
  d(X,X,s(0)). \\
  d(X{\uparrow} s(N),\, X,\, s(N){*}X{\uparrow} N ).
\\
    {\it d( F{*}G,\ X,\, F{*}D G {+} D F{*}G ) \gets  d(F,X,D F), d(G,X,D G).
      }
\end{array}
\]
\noindent
A typical query is $d(e,x,t)$, where $e,x$ are ground ($e$ represents
an expression, and $x$ a variable), $t$ is often a variable.
This may suggest moding  $d(+,+,\bot)$, however
 $d(+,\bot,\bot)$ will be sufficient for our purposes.
Consider a query {$Q=d(e_1,x_1,t_1),\ldots, d(e_n,x_n,t_n)$}
where $\seq e$ are ground.  
{\sc derivative} and $Q$ are well-3-moded under $d(+,\bot,\bot)$.
Also, the clause heads are weakly linear.
By Corollary \ref{cor.well-3-moded},
the program with $Q$ is weakly occur-check free under any selection rule.
    Alternatively,
let us find a mode $m_d$ under which the program is tidy.
The moding $m_d(2)$ of the second argument
 must be $+$, otherwise the clause body is not output linear.
As the clause heads have to be input linear, $m_d(1)=m_d(3)= -$.
Under $d(-,+,-)$ the program turns out to be tidy.
Hence, by Corollary \ref{corollary.tidy},
under any selection rule the program is occur-check free for tidy queries,
including any linear queries.
(This includes some queries not of the form considered above,
e.g. $d(X,Y,Z)$;
conversely some queries of that form are not tidy under $d(-,+,-)$,
e.g. $d(a,Y,Y)$.)
See Ex.\,\ref{ex.derivative2} for a yet another class of queries under which
the program is weakly occur-check free.

\cite{AptP94-occur-check} applied a combination of methods of well-moding and
nice moding to show that {\sc derivative} is occur-check free
for an atomic $Q$ ($n=1$) with ground $e_1,x_1$ and linear
$t_1$, under LD-resolution.
(This is subsumed by each of our two conclusions above.)
Surprisingly, 
a more general result can be obtained by a simpler approach from that work.
Under $d(-,+,-)$ {\sc derivative} is nicely-moded
and its clause heads are input linear.  
Thus it is occur-check free under LD-resolution for any nicely moded queries,
this includes any linear queries.
Still, this is subsumed by our result above employing
Corollary \ref{corollary.tidy}.

\end{example}

In this section we dealt with clause heads whose certain instances are linear.
Appendix \ref{app.weakly.tidy}
studies clauses whose certain instances are tidy,
to construct another sufficient condition for weak occur-check freeness.

\section{Final remarks}
\label{sec.final}
\subsection{Related work}

\noindent
Three approaches to assuring correctness of omitting the occur-check
have been proposed in the literature.
The first one provides syntactic sufficient conditions based on modes.
Most of the sufficient conditions considered in this paper are of this kind.
The second one employs semantic analysis of programs, mainly by abstract
interpretation
\cite{Plaisted84,Sondergaard86,CrnogoracKS96}.
The third approach is based on a correspondence between logic programs and
attribute grammars 
\cite{DeransartM85,DBLP:conf/slp/DeransartFT91,Deransart.Maluszynski93}.
A comparison (mainly of the two first approaches) is provided
by Crnogorac {\em et al.} \cite{CrnogoracKS96}

The first approach was initiated by the results on well-moded programs
\cite{AptP94-occur-check}, and nicely moded programs
\cite{AptP94-occur-check,ChadhaP94} (the latter paper uses different
terminology). 
For a well-moded (respectively nicely moded) program and query, all the
queries in LD-resolution are well-moded (nicely moded).  This, together with 
certain conditions on clause heads implies occur-check freeness
(see also \cite{apt-prolog} for a detailed presentation).
Some generalizations are presented in \cite{AptP94-occur-check}.
Generalization to other selection rules is considered in \cite{AptL95.delays};
see Section \ref{sec.tidy.condition} for comments and comparison.
Another kind of generalization of nicely moded programs
is given in \cite{ChadhaP94};
it %
considers distinct modes for various occurrences of the same predicate in
clause bodies.

We do not discuss here the other two approaches.  We only mention that
abstract interpretation usually deals with the Prolog selection rule.
Also, it may be useful to note that the notion of NSTO program in
\cite[Def.\ 2.4]{DBLP:conf/slp/DeransartFT91} (the third approach) differs
substantially from that used here, and seems rather unpractical.
One considers a set of equations between atoms, which are
taken from (renamed) clauses of the program or from the query.  
The equations correspond to all resolution steps of an SLD-derivation
(we skip the details).
The whole set is required to be NSTO.  So for instance a program
$P=\{\, p(X){\gets} q(X,X).\,\ q(Y,f(Y)).\}$ with a query $p(a)$
is considered not NSTO, as
the corresponding equation set
$\{\, p(a)\myeq p(X),\ \linebreak[3] q(X,X)\myeq q(Y,f(Y)) \,\}$   is not NSTO.
{\sloppy\par}

None of the above mentioned work attempted to weaken the NSTO condition.
An exception may be  \cite{DeransartM85}, where on p.\,139
one may find an idea similar to WNSTO introduced here.  Namely, a rather
imprecise definition of a pair $t_1,t_2$ of expressions being
 ``subject to occur check''
requires that
``occur  check is the only  reason of nonexistence of  a most general
unifier of $t_1$ and $t_2$''.  
Pair $h(X,g(X),g(X)),\ \linebreak[3] h(Y,Y,f(y))$ 
is given as an example of not being subject to occur check.
In our terms, the pair is WNSTO but not NSTO.
However the paper deals with assuring that 
each unified pair of expressions is NSTO.

In some cases, non NSTO unification of atoms can be dealt with due to
splitting such unification into a sequence of unifications of the arguments
of the atoms;
each of them may be NSTO.
For instance consider a query $Q=p(Y,Y,a)$ and a clause head
$H= p(X,f(X),X)$.  Roughly speaking, 
the unification of $Q$ and $H$ may be analysed as a sequence of unifications
of $Y\myeq X$, $Y\myeq f(X)$ and $a\myeq X$.
If abstract interpretation deals with the equations in this order,
then the need of the occur-check will apparently be established.  
If $a\myeq X$ is dealt with first, then the occur-check is found
unnecessary (as one side of each remaining equation becomes ground).

\subsection{Comments}

Let us first discuss briefly the limits of applicability of the presented
results. 
Most of the presented sufficient conditions treat predicate arguments as 
single entities, and refer
to placement of variables within certain argument positions, or to their
groundness. 
This may be not sufficient when the occur-check depends on
other features of the terms in argument positions.
For instance, in the SAT-solver of Howe and King
\cite{howe.king.tcs-shorter} 
an argument is a non-linear list of lists of pairs, 
and occur-check freeness depends on the first element of each pair being ground
\cite{Drabent.tplp18}.%
\footnote{
    For a simple %
    example, consider a program
    $P=\{ p(X,X). \}$ with a query $Q=p(f(Y,Y),f(Z,Z))$.
    It is occur-check free, however this 
    cannot be shown by the methods mentioned in this paper.
    The approach based on well-modedness does not apply, as the answer for $P$
    and $Q$ is not ground.  Treating $P$ as a nicely moded or a tidy program
    fails
    (as for $Q$ to be output linear the mode has to be $p(+,+)$,
     and this results in the clause head being not input linear).
     Similarly, the sufficient conditions employing WNSTO
     are inapplicable here.
    This should not be surprising, as $P$ with a similar query
        $Q_1=p(f(Y,g(Y)),f(Z,Z))$
    is not (weakly) occur-check free.

}   %
In such cases the methods presented or referred to here fail; instead
some analysis of the queries in SLD-trees is needed.
This is a subject of future work.
One may expect that introducing a suitable type system could be useful.

Introduction of WNSTO has two consequences. 
Some cases where unification is not NSTO can actually be safely executed
without the occur check.
Also, reasoning based on WNSTO is sometimes simpler.  For instance,
showing that program \nqueens is occur-check free was substantially more
complicated than showing it to be weakly occur-check free.
(For the former, no standard method could be used,
cf.\ Section \ref{sec.nqueens};
the latter was an immediate consequence of a syntactic sufficient condition,
cf.\ Ex.\ \ref{ex.nqueens.free.syntactic}.)
Moreover, the latter result applies to a wider class of queries.

Most of the presented sufficient conditions are based on the notion of modes.
Examples show that modes %
do not need to correspond to any intuitive 
understanding of data flow.  
(Well-moded programs are possibly an exception.)
An output argument may well be used for input data. 
Instead, the modes deal with how variables are placed in argument positions.
Neglecting this fact may be the reason why in some examples
of \cite{AptP94-occur-check} unnecessarily complicated methods were applied, 
or more general results could have been obtained.
(For
details see the comments
on  {\sc flatten} and  {\sc normalize} in Section \ref{sec.tidy.examples}, 
and on {\sc derivative} in Ex.\,\ref{ex.derivative}.)
This issue seems understood in e.g.\  \cite{ChadhaP94,CrnogoracKS96},
where one considers finding a moding under which the program (with a query)
is nicely moded.

\subsection{Conclusions}
The main contribution of this paper is weakening the notion of NSTO (not
subject to occur-check) used in the previous work on avoiding the occur-check.
Instead of requiring that no occur-check succeeds in every run of a
nondeterministic unification algorithm,
it is sufficient that one such run exists.
Additionally, we present a sufficient condition based on NSTO, generalizing 
 to arbitrary selection rules the approach based on nicely moded programs.

We generalize NSTO to WNSTO (weakly NSTO).
This leads to a generalization of the notion of occur-check free
programs/queries (based on NSTO) to {\em weakly occur-check free}
ones (based on \mbox{WNSTO}). 
We proved that unification without the occur-check is sound for any input which
is \mbox{WNSTO}. 
We presented a few sufficient conditions for WNSTO, and
for a program/query being weakly occur-check free.
Some conditions are syntactic, like   
Corollary \ref{cor.well-3-moded},
some refer to semantic notions, like Corollary \ref{corollary.weakly..free}
which explicitly refers to details of SLD-derivations.
Examples show that the proposed approach makes it possible to omit the
occur-check in some cases, to which the approaches based on NSTO are inapplicable.
In some other cases, the proposed approach leads to simpler proofs.

\DeclareRobustCommand{\lemmatidy}{\ref{lemma.tidy}}
\appendix     \section{Appendix. Proof of Lemma \lemmatidy}
\label{app.proof.tidy}

\newcommand{\commentb}[1]
{}
The proof is preceded by some
auxiliary definitions and results.

\begin{lemma}
\label{lemma.substitution}
\mbox{}
\begin{enumerate}
\item 
\label{lemma.substitution.|}
Let $\theta$ be a substitution and $S,T$ be sets of variables. %
If $S\subseteq T$ and $T\cap\Dom(\theta)\subseteq S$
 then  $\theta|S=\theta|T$.%

\item 
\label{lemma.substitution.relevant}
If $\theta$ is a relevant mgu of variable disjoint expressions $u,t$ then
$\theta|u=\theta|T$ for any $T\supseteq\Var(u)$ such that
$T\cap\Var(t)=\emptyset$.

\end{enumerate}
\end{lemma}

\noindent
{\sc Proof }
Part \ref{lemma.substitution.|} holds as 
$\theta|S\subseteq \theta|T = \theta|(T\cap\Dom(\theta))\subseteq \theta|S$.
For \ref{lemma.substitution.relevant} note that
$T\cap\Dom(\theta)\subseteq T\cap\Var(u,t)
\,\subseteq\, (T\cap\Var(u)) \cup (T\cap\Var(t)) 
\,\subseteq\, \Var(u) \cup\emptyset$.
Now by~\ref{lemma.substitution.|}, $\theta|Var(u)=\theta|T$.
\hfill$\Box$

\begin{definition} %
Let $\S$ be a set of variables.
A substitution
$\theta$ is {\bf linear for} $\S$ if for any two distinct variables 
$X,Y\in {\S}$ the pair $X\theta,Y\theta$ is linear.
If ${\S}=\{X\}$ is a singleton then we require that $X\theta$ is linear.

$\theta$ is linear for an expression
$ t$ if $\theta$ is linear for $\Var({ t})$.

\end{definition}
A  $\theta$ linear for $\Dom(\theta)$ is linear in the sense of 
\cite[Def.\,A.3]{AptP94-occur-check}
(this means that if $\theta=\{X_1/t_1,\ldots,X_n/t_n\}$ then
$\seq t$ is linear).
Obviously, if $\theta$ is linear for a set
${ S}$ then it is linear for any  ${ S}'\subseteq{ S}$.  Also:

\begin{remark}
\label{cor.apply.linear} 
\mbox{}\ 
If a substitution $\theta$ is linear for $\{X,Y\}$, where $X,Y$ are distinct
variables then \linebreak[4]$\Var(X\theta)\cap\Var(Y\theta)=\emptyset$.  
\sloppy
\end{remark}

\begin{lemma}
\label{lemma.apply.linear}
If a substitution $\theta$ is linear for a linear expression ${ t}$ then
$t\theta$ is linear.
\end{lemma}
\noindent
{\sc Proof }
Let $\theta|t=\{X_1/u_1,\ldots,X_n/u_n\}$ and
$\Var(t)\setminus\Dom(\theta)= \{X_{n+1},\ldots,X_m\}$.  Let $u_i=X_i$ for
  $i\in\{n{+}1,\ldots,m\}$.
So $X_i\theta=u_i$ (for $i=\{1,\ldots,m\}$), hence $\seq[m]u$ is linear.
Consider an occurrence of a variable $V$ in $t\theta$.  We have two
possibilities.  1.\,\,The occurrence is within a subterm $u_i$ introduced by
$\theta$;
$V\in\Var(u_i)$ (for a unique $i\in\{1,\ldots,n\}$), hence $V\in\Var(X_i\theta)$.
2.\,\,Otherwise the occurrence is within $t$; $V\in\Var(t)\setminus\Dom(\theta)$ and
$V=X_i=X_i\theta$ (for a unique $i\in\{n+1,\ldots,m\}$).
{\sloppy\par}

So $V$ occurs exactly once in exactly one $u_i$. 
Assume that $V$ occurs twice in $t\theta$.  Then it occurs in two occurrences
of some $u_i$ in $t\theta$. %
Hence $X_i$ occurs twice in $t$. Contradiction. 
\hfill$\Box$

\medskip

\begin{lemma}\rm
\label{lemma.3variables}
Let $\theta$ be a substitution and $X,V,V'$ variables.  Assume that
$X\in\Var(V\theta)$ and $X\in\Var(V'\theta)$.  Then $X=V=V'$, or 
$X,V,V'\in\Var(\theta)$ and moreover $X\in\Ran(\theta)$.
\end{lemma}

\noindent
{\sc Proof }
Assume $X\not\in\Ran(\theta)$.  
From $X\in\Var(V\theta)\subseteq \{V\}\cup\Ran(\theta)$ it follows  $X=V$.
In the same way we obtain $X=V'$.

Assume $V\not\in\Var(\theta)$.  Hence $V\theta=V$, thus $X=V$. Now
$V\in\Var(V'\theta) \subseteq \{V'\}\cup \Var(\theta)$ implies $V=V'$.  
In the same way,  $V'\not\in\Var(\theta)$ implies  $X=V=V'$. \hfill$\Box$

\begin{lemma} %
\label{lemma.likeA5}
Consider two variable disjoint expressions $s,t$ where $t$ is linear.  
  Let $ \S$ be a set of variables such that $\S\cap\Var(t)=\emptyset $.
If $s,t$ are unifiable then there exists a relevant  %
mgu $\theta$ of $s,t$ such that
\\\hspace*{2em}
  \begin{tabular}[t]{l}
    $\theta$ is linear for $\S$, and 
\\
    $\Ran(\theta|s) \subseteq \Var(t)$.
  \end{tabular}
\end{lemma}

This lemma is similar to \cite[Lemma A.5]{AptP94-occur-check}, attributed
to \cite{DembinskiM85}.
The main difference is that there $\S=\Var(s)$ and one deals with $\theta|s$
instead of $\theta$.

\medskip
\noindent
{\sc Proof }
By structural induction on $t$.  Without loss of generality, we can assume that
$s,t$ are terms, and that a function symbol has at most two arguments
(as the set of unifiers of $f(\seq s)$ and $f(\seq t)$ is the same as that for
$g(s_1,g(s_2,\ldots g(s_{n-1},s_n)\cdots)) $
and $g(t_1,g(t_2,\ldots g(t_{n-1},t_n)\cdots)) $).
We can assume that $ S$ contains at least two
variables
 (otherwise a variable not occurring elsewhere may be added to $ S$).

Assume $s,t$ are unifiable.  
We have the following five cases.

1. $s$ is a variable. 
Let  $\theta=\{s/t\}$.  Obviously,
\commentb{
}
$\Ran(\theta|s)\subseteq \Var(t)$.
Consider distinct $X_1,X_2\in\S$.
Note that expression $X_1,X_2,t$ is linear.
For $i=1,2$, 
either  $X_i\theta=X_i$ or  $X_i\theta=t$;
the latter holds for at most one of  $X_1,X_2$.  Hence $X_1\theta, X_2\theta$
is linear.

2. $t$ is a variable. The claim holds for $\theta=\{t/s\}$, as
$\theta|s=\epsilon$ (the empty substitution).

3. $s=t$ is a constant.  The claim holds for $\theta=\epsilon$.

4. $s=f(s_1)$ and $t=f(t_1)$.  Assume that the claim holds for $s_1,t_1$.
Then it obviously holds for $s,t$. %

5. $s=f(s_1,s_2)$ and $t=f(t_1,t_2)$.
We split the unification of $s,t$ in two steps.
As  $s,t$ are unifiable, $s_1,t_1$ are.
By the inductive assumption,
there exists a relevant mgu $\sigma$ of $s_1$ and $t_1$ such that 
\commentb{}
$\sigma$ 
is linear for $\S$ \commentb{} and 
$\Ran(\sigma|s_1)\subseteq\Var(t_1)$. 
Note that $t_2\sigma=t_2$
(as   $t_2$ is variable disjoint with $t_1,s_1$).
Now by \cite[Lemma 2.24]{apt-prolog},  $s_2\sigma$ and $t_2$ are unifiable,
and if $\rho$ is their mgu then $\sigma\rho$ is an mgu of $s$ and $t$.

By the inductive assumption,
there exists a relevant mgu $\rho$  of $s_2\sigma$ and $t_2$ such that 
\commentb{}
$\rho$ is linear for $\S\cup\Var(t_1)$ and
$Ran(\rho|s_2\sigma)\subseteq\Var(t_2)$.
So the mgu $\theta=\sigma\rho$ of $s$ and $t$ is relevant
(as $\Var(\theta)\subseteq\Var(\sigma)\cup\Var(\rho)\subseteq\Var(s,t)$).

 To show that $\theta$ is linear for \S, 
we consider distinct variables $X_1,X_2\in\S$ %
and show that  $(X_1,X_2)\sigma\rho$ is linear.
First %
$(X_1,X_2)\sigma$ is linear, as $\sigma$ is linear for \S.
Note that (for $i=1,2$) $X_i\neq X_i\sigma$ implies
 $X_i\in\Var(\sigma)\cap\S\subseteq \Var(s_1)$,
and then $\Var(X_i\sigma)\subseteq\Ran(\sigma|s_1)$.
Hence
$Var((X_1,X_2)\sigma)\subseteq\{X_1,X_2\}\cup\Ran(\sigma|s_1)\subseteq\S\cup\Var(t_1)$.
Now, as $\rho$ is linear for $\S\cup\Var(t_1)$, 
$(X_1,X_2)\sigma\rho$ is linear.

It remains to show that  $\Ran(\theta|s) \subseteq \Var(t)$.
We do this by showing that if $X/w\in\theta|s$ then $Vars(w)\subseteq\Var(t)$.
Consider a pair $e\in\theta|s$.  
As $\theta=\sigma\rho$, by the definition of composition of substitutions
(cf.\ \cite[Lemma 2.3]{apt-prolog})
we have two cases.

i. $e$ is $X/u\rho$ where $X/u\in\sigma|s=\sigma|s_1$.
Hence $Var(u)\in\Ran(\sigma|s_1)\subseteq\Var(t_1)$.
By Lemma   \ref*{lemma.substitution}.\ref{lemma.substitution.relevant},
 $\rho|s_2\sigma=\rho|(s_2\sigma,t_1)$ (as $\Var(t_1)\cap\Var(t_2)=\emptyset$).
Thus
$u\rho = u(\rho|s_2\sigma)$.
Hence
\commentb{}
$\Var(u\rho)\subseteq\Var(u)\cup\Ran(\rho|s_2\sigma)\subseteq
 \Var(t_1)\cup\Var(t_2)=\Var(t)$.

ii. $e$ is $X/u\in\rho$ where $X\in\Var(s)$.  So $X\not\in\Var(t_2)$.
As $\Var(\rho)\in\Var(s_2\sigma,t_2)$, it follows that 
$X\in\Var(s_2\sigma)$.
Thus $\Var(u)\subseteq\Ran(\rho|s_2\sigma)\subseteq\Var(t_2)$. 
\hfill $\Box$

\medskip
We are ready for the main proof of this section.

\smallskip\smallskip
\noindent
{\bf Lemma \ref{lemma.tidy}}
{\ \it 
  Let $Q$ be a tidy query, and $C$ a tidy clause
  variable disjoint from $Q$.
  An SLD-resolvent $Q'$ of $Q$ and $C$ is tidy.
}

\smallskip
\noindent
{\sc Proof }
Let   $Q=A_1,\vec A $  where $\vec A =A_2,\ldots,A_n$ and 
$C = H\gets\vec B$, where $\vec B=\seq[m]B$.
As the order of atoms in queries does not matter,
we may assume that $A_1$ is the selected atom of $Q$.
Let 
\[
\mbox{ $A_1=p(s_1;{\dred t_1})$, \qquad\qquad\qquad   $H=p({\dred s};t)$.}
\]
$A_1$ and $H$ are variable disjoint (as $Q$ and $C$ are).
As $Q$ and $C$ are tidy, $A_1$ is input-output disjoint, and $t_1$ and $s$
are linear.  Consider
\[
\mbox{$\sigma$  --   a relevant mgu of $s_1$ and $s$, \qquad and \qquad
  $\rho$ -- a relevant mgu of $t_1$ and $t\sigma$.
}
\]
Note that $t_1=t_1\sigma$, as no variable from $t_1$ occurs in $s_1,s$, hence
in $\sigma$.  So $\theta=\sigma\rho$ is a relevant mgu of $A_1$ and $H$
(mgu by \cite[Lemma 2.24]{apt-prolog}, relevant by
$\Var(\theta)\subseteq\Var(\sigma)\cup\Var(\rho)\subseteq\Var(A_1,H)$).

Let $S_{o B}$ be the set of variables in the output positions of $\vec B$.
As $C$ is tidy, $S_{o B}\cap\Var(s)=\emptyset$.
By Lemma~\ref{lemma.likeA5}, $\sigma$ may be chosen so that 
\[
\mbox{
\commentb{} $\sigma$ is linear for $\Var(Q)\cup S_{o B}$ \ \
and \ \ $\Ran(\sigma|s_1)\subseteq\Var(s)$.
}
\]
By Lemma \ref{lemma.apply.linear},
 $(\vec B,Q)\sigma$ is output linear (as $Q,\vec B$ is).
From $\Var(t_1)\cap\Var(s_1,s,t,\vec B)=\emptyset$ it follows that
$t_1$ and $(s_1,t,\vec B)\sigma$ 
 are variable disjoint.
Let $t_i$ be the terms on the output positions of $A_i$, for $i=2,\ldots,n$.
Let $S_{o Q} = \Var(t_2\sigma,\ldots,t_n\sigma)$.
As $Q\sigma$ is output linear, $\Var(t_1)\cap S_{o Q}=\emptyset$.
By Lemma~\ref{lemma.likeA5},
the mgu $\rho$ of $t_1$ and $t\sigma$ can be chosen so that
\[
\mbox{
  \commentb{}
  $\rho$is linear for $\Var(s_1\sigma,t\sigma,\vec B\sigma)\cup S_{o Q}$
  \ \ and \ \ 
$\Ran(\rho|t\sigma)\subseteq\Var(t_1)$.
}
\]
\comment{
}
\noindent
Hence by Lemma \ref{lemma.apply.linear} %
\[
Q' = (\vec B,\vec A)\sigma\rho
\qquad \mbox{ is output linear}
\vspace{-2ex}
\enlargethispage{2.4ex}
\]%
(as $(\vec B,Q)\sigma$ is).
\noindent

We proved that the SLD-resolvent $Q'$ of $Q$ and $C$ is output linear.
It remains to show that $\to_{Q'}$ is acyclic.  We consider all atom pairs
$L\theta\to_{Q'}L'\theta$ from $Q'$
 (where $L,L'$ are from $\vec B,\vec A$),
 and relate them to relations $\to_{Q}$
and $\to_{\vec B}$\,.
For each such pair there exists a variable
$X\in\VarOut(L\theta)\cap\VarIn(L'\theta)$.
So there exist variables $V,V'$, and then $W,W'$
such that
\[
\label{eq.VYYWW}
    \begin{array}{l l l l}
    V\in\VarOut(L\sigma), &
    V'\in\VarIn(L'\sigma), &  X\in\Var(V\rho), & X\in\Var(V'\rho),
    \\
    W\in\VarOut(L), & W'\in\VarIn(L'), &
    V\in\Var(W\sigma), & V'\in\Var(W'\sigma),
    \end{array}
\ \ (\mbox{where } \ \ L\theta\to_{Q'}L'\theta).
\]
(The variables may be not distinct.)
Note that, as $Q'$ is output linear,
$X$ occurs only once in the output positions of $Q'$, hence $V$ is unique.
As $(\vec B,\vec A)\sigma$
 is output linear,
$V$ occurs only once in the output positions of
 $(\vec B,\vec A)\sigma$, hence $W$ is unique.
As $\vec B,\vec A$ is output linear, $W$ occurs only once in its
output positions.

We consider various cases of $L\theta\to_{Q'}L'\theta$.

\smallskip
\noindent
1.  $L=B_i$  (for some $i\in\{1,\ldots,m\}$).  %
As $\VarOut(B_i)\cap\Var(s_1,s)=\emptyset$ and
$\Var(\sigma)\subseteq\Var(s_1,s)$, we have $W\not\in\Var(\sigma)$, hence
$W\sigma=W$.  Thus $W=V$.
\medskip\noindent
1a.  $L=B_i$ and  $L'=B_j$  (for some $j\in\{1,\ldots,m\}$).
As  $\rho$ is linear for $\Var(\vec B\sigma)$
(hence for $\{V,V'\}$), and $\Var(V\rho)\cap\Var(V'\rho)\not=\emptyset$,
by Remark \ref{cor.apply.linear} we obtain $V=V'$.

As  $V'\in\VarIn(B_j\sigma)\subseteq\VarIn(B_j)\cup\Var(\sigma)$ and
 $V'=W\not\in\Var(\sigma)$,
we have $W\in\VarIn(B_j)$.  %
  So $B_i\to_{Q_{B}}\!B_j$
(as $W\in\VarOut(B_i)$).

\medskip\noindent
1b.   $L=B_i$ and  $L'=A_j$  (for some $j\in\{2,\ldots,n\}$).
Note that $W\not\in\Var(Q)$ and $W\not\in\Var(\sigma)$ implies
$W\not\in\VarIn(A_j\sigma)$. 
As $V'\in\VarIn(A_j\sigma)$, we obtain $V'\neq W$.
So by Lemma \ref{lemma.3variables},
$X,W,V'\in\Var(\rho)\subseteq\Var(t_1,t\sigma)$.
Thus
 \mbox{$W\in\Var(t\sigma)$} (as $W\not\in\Var(Q)$).

Assume $V'\in\Var(t\sigma)$. 
As $X\in\Var(W\rho)\cap\Var(V'\rho)$
and $\rho$ is linear for $t\sigma$,
we have contradiction by Remark \ref{cor.apply.linear}.
Thus  $V'\in\Var(t_1)$, hence $V'\sigma=V'$. 

Note that $V'\in\Var(V'\sigma)\cap\Var(W'\sigma$),
\commentb{and  $Q\sigma= $.}
$V',W'\in\Var(Q)$ and
  $\sigma$ is linear for $\Var(Q)$.  %
Thus $V'=W'$ by Remark \ref{cor.apply.linear}.
So $W'\in\Var(t_1)=\VarOut(A_1)$, hence  $A_1\to_Q A_j$
(as $W'\in\VarIn(A_j)$).

\medskip\noindent
2. $L=A_i$ (for some $i\in\{2,\ldots,n\}$).
As $Q$ is output linear, $W\not\in\Var(t_1)$.
We first show that $W\neq W'$ implies $W'\in\Var(A_1)$ and  $W\in\Var(s_1)$,
and thus %
$A_i\to_Q A_1$ (as $W\in\VarOut(A_i)$).

Assume  $W\neq W'$. Note that $X\in\Var(W\theta)\cap\Var(W'\theta)$. 
By  Lemma \ref{lemma.3variables}, $X,W,W'\in\Var(\theta)\subseteq\Var(A_1,H$).
As $W,W'\not\in\Var(H)$, $W,W'\in\Var(A_1)$.  As $W\not\in\Var(t_1)$, 
$W\in\Var(s_1)$.

\medskip\noindent
2a. \ $L=A_i$ and $L'=B_j$ (for some $j\in\{1,\ldots,m\}$).
As  $W\neq W'$, we obtain $A_i\to_Q A_1$.

\medskip\noindent
2b. $L=A_i$ and $L'=A_j$ (for some $j\in\{2,\ldots,n\}$).
Employing $W$ and $W'$ we are going to show that
$A_i\to_Q^+ A_j$. 
If $W=W'$ the claim is obvious (as $A_i\to_Q A_j$).
So assume $W\neq W'$.
Hence $A_i\to_Q A_1$.  Also, 
 $W'\in\Var(A_1)$ and  $W\in\Var(s_1)$.
Thus $V'\in\Var(A_1\sigma)$ and  $V\in\Var(s_1\sigma)$.

As $\sigma$ is linear for $\Var(Q)$, 
by Lemma \ref{lemma.apply.linear} $(W,W')\sigma$ is linear. 
Thus $V\neq V'$.
Assume $V,V'\in\Var(s_1\sigma)$.  As 
$\rho$ is linear for $\Var(s_1\sigma)$,
by Remark \ref{cor.apply.linear} we obtain contradiction
with $X\in\Var(V\rho)\cap\Var(V'\rho)$.
As $V\in\Var(s_1\sigma)$ and $V'\in\Var(A_1\sigma)$, we have 
$V'\in\Var(t_1\sigma)=\Var(t_1)$, and thus $V'\sigma=V'$.

From $V'\sigma=V'$,
exactly as in case 1b,\ it follows that $W'=V'$, and then $A_1\to_Q A_j$.

So we showed that  $A_i\theta\to_{Q'}A_j\theta$ implies
$A_i\to_Q^l A_j$ where $l=1$ or $l=2$.

\medskip %
Now assume that $\to_{Q'}$ has a cycle $\cal C$.
Suppose that $\cal C$ contains only atoms from $Q\theta$, say
\linebreak[4]
$A_{i_1}\!\theta\to_{Q'}\cdots\to_{Q'}A_{i_k}\!\theta\to_{Q'}A_{i_1}\!\theta$
(for some $k,\seq[k]i\in\{1,\ldots,n\}$).  
Hence  $A_{i_1}\to_{Q}^+\cdots\to_{Q}^+A_{i_k}\to_{Q}^+A_{i_1}$;
relation $\to_Q$ has a cycle. Contradiction, as $Q$ is tidy.
In the same way it follows that $\cal C$ cannot contain only atoms from
$\vec B\theta$
(as otherwise $\to_{\vec B}$ has a cycle, contradiction with $H\gets \vec B$ being
tidy). 
So $\cal C$ contains atoms from $Q\theta$ and from $\vec B\theta$.  Hence there
exist in  
$\cal C$ atoms $B_j\theta$ and  $B_{k}\theta$ ($j,k\in\{1,\ldots,m\}$) such
that all the atoms 
between them in $\cal C$ are from $Q\theta$, and there exists at least one
such atom: \
$B_j\theta\to_{Q'}A_i\theta\to_{Q'}^*A_{i'}\theta\to_{Q'}B_k\theta$
(where $i,i'\in\{1,\ldots,n\}$).
Hence
$A_1\to_{Q}A_i\to_{Q}^*A_{i'}\to_{Q}A_1$; %
relation $\to_Q$ contains a cycle.
Contradiction.  So $\to_{Q'}$ is acyclic.  Hence $Q'$ (previously shown to be
output linear) is tidy.
{\sloppy\par}

Any resolvent $Q''$ of $Q$ (with selected $A_1$) and a variant of $H\gets \vec B$
is a variant of $Q'$, hence $Q''$ is tidy.
\hfill  $\Box$

\DeclareRobustCommand{\lemmaiteration}{\ref{lemma.iteration}}
\section{Appendix. Proof of Lemma \lemmaiteration}
\label{app.proof.lemma.iteration}

The proof employs a technical lemma.
\begin{lemma}
  \label{lemma.MMA.mgu}
  Consider a run $R$ of MMA producing an mgu $\theta$.  Let 
  $X_1\myeq t_1,\ldots,X_k\myeq t_k$ (in this order) be the equations selected
  in action (5) in $R$, and $\gamma_i=\{X_i/ t_i\}$ for $i=1,\ldots,k$.  Then
   $\theta=\gamma_1\cdots\gamma_k$.  Moreover, the last equation set of $R$ is 
$E=\{X_i\myeq t_i\gamma_{i+1}\cdots\gamma_k \mid 0<i\leq k \,\}$.
\end{lemma}
{\sc Proof }
Note that $\seq[k]X$ are distinct.
Action (5) means applying $\gamma_i$ to all the equations except for 
$X_i\myeq t_i$.  Moreover, (a current instance of) $X_i\myeq t_i$ is not
selected anymore in $R$.  So $E$ contains 
$X_i\myeq t_i\gamma_{i+1}\cdots\gamma_k$, hence
$\theta$ contains %
$X_i/ t_i\gamma_{i+1}\cdots\gamma_k$ (for $i=1,\ldots,k$).
Let $\psi_j=\{X_i/t_i\gamma_{i+1}\cdots\gamma_j \mid 0<i\leq j \,\}$ and
$\varphi_j=\gamma_{1}\cdots\gamma_j$, for $j=0,\ldots,k$.
Now $\psi_j=\varphi_j$, by induction on $j$ 
(as
$
\{X_i/t_i\gamma_{i+1}\cdots\gamma_j \mid 0<i\leq j \,\}\gamma_{j+1}=
\{X_i/t_i\gamma_{i+1}\cdots\gamma_{j+1} \mid 0<i\leq j \,\}\cup\gamma_{j+1}
$).
Thus $\theta=\varphi_k$.
\hfill$\Box$

\smallskip
\noindent
{\bf Lemma \ref{lemma.iteration}} \ %
{\it  Let $E_1\cup E_2$ be an equation set and $E_1$ be WNSTO.

  If $E_1$ is not unifiable then  $E_1\cup E_2$ is WNSTO.

  If  $\theta_1$ is an mgu of $E_1$, and $E_2\theta_1$ is WNSTO
  then  $E_1\cup E_2$ is WNSTO.
} %
\smallskip
\noindent

\noindent
{\sc Proof }\ %
Assume that $E_1$ is not unifiable.  
Then from an occur-check free run $R$ of MMA on  
$E_1$ one can construct in an obvious way an occur-check free run $R'$ of MMA on 
$E_1\cup E_2$, performing the same actions and selecting the same equations.

Now assume that $\theta_1$ is an mgu of $E_1$.
Consider an occur-check free run $R_1$ of MMA on $E_1$, and an occur-check
free run $R_2$ of MMA on $E_2\theta_1$. 
Without loss of generality we may assume that $\theta_1$ is the result of
$R_1$.  (Otherwise, $R_1$ produces $\theta$ such that
$E_2\theta_1$ is a variant of $E_2\theta$.)
\commenta{
 A better justification.  Some lemma from Apt ?
}
Let step (5) be applied in $R_1$ to equations 
$X_1\myeq t_1,\ldots,X_k\myeq t_k$ (in this order), 
and in  $R_2$ to 
$X_{k+1}\myeq t_{k+1},\ldots,X_m\myeq t_m$ (in this order).
Let $\gamma_i=\{X_i/t_i\}$, for $i=1,\ldots,m$.
All $X_i$ ($i=1,\ldots,m$) are distinct (as for $i=1,\ldots,k$
no $X_i$ occurs in $E_2\theta$).
Let $F$ be the last equation set in $R_1$;
by Lemma \ref{lemma.MMA.mgu},
$F=\{X_1\myeq t_1\gamma_2\cdots\gamma_k,\ldots,X_k\myeq t_k\}$.
{\sloppy\par}

To construct a run $R$ of MMA on $E_1\cup E_2$, 
for each equation set from $R_1$ or $R_2$ we construct a corresponding
equation set from $R$.

Consider an equation set $E$ in $R_1$.
Let $X_1\myeq t_1,\ldots,X_i\myeq t_i$ be the equations on which 
action (5) has been performed in $R_1$ until obtaining $E$
($i\in\{0,\ldots,k\}$). 
The equation set corresponding to $E$ is
$E' = E\cup E_2\gamma_1\cdots\gamma_i$.
In particular (by Lemma \ref{lemma.MMA.mgu}),
$F\cup E_2\theta_1$ corresponds to the last equation $F$ of $R_1$.

Consider an equation set $E$ in $R_2$.
Let $X_{k+1}\myeq t_{k+1},\ldots,X_i\myeq t_i$ %
be the equations on which 
action (5) has been performed in $R_2$ until obtaining $E$
($i\in\{k,\ldots,m\}$). 
Let the corresponding equation set in $R$ be
$E'=F\gamma_{k+1}\cdots\gamma_i\cup E$.

Consider now the sequence consisting of the equation sets corresponding to
those of
$R_1$ and then of the equation sets corresponding to those of $R_2$ (without
the first one, to avoid a repetition of $F\cup E_2\theta_1$).
The sequence is a run of MMA on $E_1\cup E_2$.
As the run does not involve action (6), $E_1\cup E_2$ is WNSTO.
\hfill$\Box$

\medskip

\section{Appendix. Another syntactic sufficient condition}
\label{app.weakly.tidy}

\noindent
Here we present a sufficient condition for avoiding the
occur-check, it employs WNSTO and is
based on the syntactic conditions for tidy programs 
(Corollary \ref{corollary.tidy}).

Consider a 3-moding $M$ and an additional moding $M'$,
for the latter we use symbols $+',-'$.
Consider transforming each clause $C$ of a program $P$, by grounding the
variables 
that occur in the $+$ positions in the head.
Let $P'$ be the resulting program.
Let us say that $P$ is {\bf weakly tidy} (under $M,M'$)
 if $P'$ is tidy under $M'$.

For an example, let $P$ be a program containing a clause $C$ with body
$B = q(X,Y),\linebreak[3] q(Y,Z),\linebreak[3] q(Z,X)$.
Assume also that $P$ contains a clause head $H=q(t,u)$ with 
$t,u$ containing a common variable.
For $P$ to be tidy, the argument positions of $q$ cannot be both input (due
to $H$), and cannot be both output (due to $B$).
However, if they are $(+,-)$, or $(-,+)$ then $\to_B$ is cyclic.
Thus $P$ is not tidy (and not nicely moded with input linear heads)
under any moding.
Assume now that $C$ is $p(X)\gets B$.
Under $p(+)$, and $q(+',-')$
the clause is weakly tidy.  (A corresponding clause of $P'$ is
$ p(s) \gets q(s,Y),\linebreak[3] q(Y,Z),\linebreak[3] q(Z,s)$, for a
ground term $s$; note that $q(+',-')$ may be replaced by $q(-',+')$.)
By the lemma below,
if $p(s)$ is selected in a tidy query $Q$ then the resolvent of $Q$ and $C$
is tidy.

\begin{lemma}
\label{lemma.weakly.tidy}

Let $P$ be a weakly tidy program under $M,M'$, and $Q$ a query tidy under $M'$.
Under any selection rule compatible with $M$

each query in any SLD-derivation for $P$ with $Q$ is tidy under $M'$, and

$P$ with $Q$ is weakly occur-check free.

\end{lemma}

\noindent
{\sc Proof }
In the proof, by tidy we mean tidy under $M'$.
Let $A$ be the selected atom of a tidy %
 query $Q$, and $H$ be the head of
a clause $C$ of $P$ variable disjoint from $Q$.
Let $A=p(s;t;u)$ and $H=p(s';t';u')$ (under $M$).
Unifying $A$ with $H$ can be divided in two steps:

\  1.\ Unifying  $s$ with $s'$.
   As $s$ is ground, $s\MYEQ s'$ is NSTO.

\  2.\ Unifying $(t,u\MYEQ t',u')\theta$ 
provided that $s\MYEQ s'$ is unifiable with an mgu $\theta$.
This is similar to unifying $A\theta\myeq H\theta$;
in both cases the unifiers are the same.
Note that $Q\theta=Q$ (thus $A\theta=A$), $C\theta$ is tidy,
 and $\Var(C\theta)\subseteq\Var(C)$.
So $A\theta$ is an atom from a tidy query, and $H\theta$
is the head of a tidy clause $C\theta$ variable disjoint from $Q$.
By Lemma \ref{lemma.tidy}, any resolvent $Q'$ of 
$Q\theta$ and $C\theta$ is tidy.
By Corollary \ref{corollary.tidy}, $A\theta\myeq H\theta$ is NSTO.
Hence  $(t,u\MYEQ t',u')\theta$ is WNSTO.
(This is because if $(e,e')\MYEQ(e,e'')$ is NSTO then $e'\MYEQ e''$ is WNSTO,
for any expressions $e,e',e''$; we skip details of a proof based on 
constructing an appropriate run of MMA out of a run of MMA on 
 $(e,e')\MYEQ(e,e'')$ which first deals with  $e\MYEQ e$.)

Out of 1.\ and 2.\ 
by Corollary \ref{corollary.iteration} it follows that $A\myeq H$ is WNSTO.
If $A\myeq H$ is unifiable then
any resolvent $Q'$ of $Q\theta$ and $C\theta$ is also a resolvent of 
$Q$ and $C$ (as by \cite[Lemma 2.24]{apt-prolog}
the mgu $\theta$ of  $s\MYEQ s'$ exists, $(t,u\MYEQ t',u')\theta$ 
is unifiable, and if $\sigma$ is its mgu then
$\theta\sigma$ is an mgu of $A$ and $H$).
As any resolvent $Q''$ of $Q$ and $C$ is a variant of $Q'$, 
the resolvent is tidy.

We showed, for a tidy (under $M'$) query $Q$ and a standardized apart clause $C$ of $P$,
that unification of the selected atom of $Q$ with the head of $C$ is WNSTO,
and that the resolvent (if it exists) of $Q$ with $C$ is tidy.
By simple induction, the same holds for each query in any derivation
for $P$ with $Q$ (under a selection rule compatible with $M$).
\hfill$\Box$
\medskip

For well-3-moded program and queries, certain $+$ positions of the queries in
SLD-trees are guaranteed to be ground.  
By Lemmas \ref{lemma.well-3-moded} and \ref{lemma.weakly.tidy} we obtain the
following property.

\begin{corollary}
\label{cor.weakly.tidy}
Let $P$ be a weakly tidy program under $M,M'$, and $Q$ a query tidy under $M'$.
Let $P$ and $Q$ be well-3-moded under $M$.

$P$ with $Q$ is weakly occur-check free under the Prolog selection rule.

If no argument position is moded by $M$ as $-$ %
then $P$ with $Q$ is weakly occur-check free under any selection rule.
  
\end{corollary}

\begin{example}\rm
  \label{ex.derivative2}
  Program {\sc derivative} is weakly occur-check
  free for some queries not considered in Ex.\,\ref{ex.derivative}.
  Consider modings $M=d(\bot,+,\bot)$ and 
  $M' = d(+',\bullet,-')$, where $\bullet$ stands for $+'$ of $-'$.
  {\sc derivative} is weakly tidy under $M,M'$ (as 
  {\sc derivative} with $X$ replaced by a constant is tidy under $M'$).
  Assume an initial query $Q=d(e_1,x_1,t_1),\ldots, d(e_n,x_n,t_n)$,
  where terms $\seq x$ are ground.
  {\sc derivative} with $Q$ is well-3-moded under $M$.
  By  Corollary \ref{cor.weakly.tidy}, if $Q$ is tidy under $M'$ then
  {\sc derivative} with $Q$ is weakly occur-check free, under any selection
  rule.

  An example of such tidy query is any $Q=d(e,x,t)$, where $t$ is linear and
  variable disjoint with $e$ (and $x$ is ground).
  Note that when $e$ is not linear then such query is not dealt with by
  the sufficient conditions discussed in Ex.\,\ref{ex.derivative}.

\end{example}

\bibliographystyle{eptcsalpha}
\bibliography{bibpearl,bibmagic,bibs-s,bibshorter,ja}

\commenta
{%
{\bf A note for reviewers}. \cite{apt-prolog} is available at
\small
\url{https://homepages.cwi.nl/~apt/book.ps}.
}
\end{document}